# Techno-economic model of a second-life energy storage system for utility-scale solar power considering li-ion calendar and cycle aging

Ian Mathews[1,*], Bolun Xu[2], Wei He[1], Vanessa Barreto[3], Tonio Buonassisi[1] and Ian Marius Peters[1]

[1]*Department of Mechanical Engineering, Massachusetts Institute of Technology*
[2]*MIT Energy Initiative, Massachusetts Institute of Technology*
[3]*Sloan School of Management, Massachusetts Institute of Technology*

*Correspondence: imathews@mit.edu

**Abstract**

While the use of energy storage combined with grid-scale photovoltaic power plants continues to grow, given current lithium-ion battery prices, there remains uncertainty about the profitability of these solar-plus-storage projects. At the same time, the rapid proliferation of electric vehicles is creating a fleet of millions of lithium-ion batteries that will be deemed unsuitable for the transportation industry once they reach 80% of their original capacity. The repurposing and deployment of these batteries as stationary energy storage provides an opportunity to reduce the cost of solar-plus-storage systems, if the economics can be proven.

We present a techno-economic model of a solar-plus-second-life energy storage project in California, including a data-based model of lithium nickel manganese cobalt oxide battery degradation, to predict its capacity fade over time, and compare it to a project that uses a new lithium-ion battery. By setting certain control policy limits, to minimize cycle aging, we show that a system with SOC limits in a 65–15% range, extends the project life to over 16 years, assuming a battery reaches its end-of-life at 60% of its original capacity. Under these conditions, a second-life project is more economically favorable than a project that uses a new battery and 85–20% SOC limits, for second-life battery costs that are <80% of the new battery. The same system reaches break-even and profitability for second-life battery costs that are <60% of the new battery. In comparison, control policies that more rapidly degrade the battery by using a large 80% DoD limit in a 95–15% range, to 'front-load' revenue during a project life-cycle, were found to significantly underperform a new battery owing to the resulting shorter project life.

Our model shows that using current benchmarked data for the capital and O&M costs of solar-plus-storage systems, and a semi-empirical data-based degradation model, it is possible for EV manufacturers to sell second-life batteries for <60% of their original price to developers of profitable solar-plus-storage projects.

**Introduction**

The use of lithium-ion batteries to minimize the impact of the variability of supply from renewable power sources on the grid is growing rapidly. Utility-scale stationary energy storage systems are being deployed to store excess energy when production exceeds demand and minimize curtailment in markets with high storage penetration, such as California [1]. While the benefits of the increased use of energy storage in combination with renewable energy generation are clear, the economics for project developers are still uncertain. There remains a need for newer- and lower-cost technologies and hardware to enable the widespread use of stationary energy storage systems, but also a need for data-based modeling and intelligent controls to extract as much value as possible from the technologies that do exist [2], [3].

At the same time as the need for stationary energy storage systems is growing, the rapid proliferation of electric vehicles (EV) is creating a fleet of millions of lithium-ion batteries that will be deemed unsuitable for the rigorous transportation duty cycle/environment after a number of years operation. It is expected that EV owners will replace their battery system once they have lost just 20% of their capacity [4]. These used batteries present a massive opportunity to be repurposed for new applications where the duty cycling and current levels are less onerous than EV driving — potentially providing a low-cost source of lithium-ion batteries for new applications, increasing a battery's lifetime value and postponing the eventual cost of recycling [5]. Multiple GWhs of these so called 'second-life batteries' are expected to become available in the coming years [6].

Combining second-life batteries with grid-scale solar energy systems is potentially a good application for these EV batteries because the energy and power requirements will be moderate compared to propulsion specifications with multiple small-scale pilot projects now demonstrated [7], [8]. However, a key hurdle to deploying second-life batteries at scale in large power infrastructure projects is establishing the expected performance and lifetime in their second use.

Recently, there has been a growth in the number of methods available to predict the capacity fade of lithium-ion batteries based upon measured degradation data including; semi-empirical models that tune algorithms to specific chemistries [9], probabilistic methods such as gaussian processes that achieve high accuracy but require larger datasets [10], and other machine learning approaches using neural networks that again require large datasets [11]. Xu et al. [9] proposed a semi-empirical non-linear model that was trained using lab-measured degradation behavior and considers the effects of solid-electrolyte-interface (SEI) film formation, calendar aging, cycle aging and temperature for a number of lithium-ion battery chemistries including nickel-manganese-cobalt (NMC) typically used in EVs.

In this paper, we model the economic performance of a combined photovoltaics plus second-life energy storage project in California using current capital cost benchmarks [12], and use this data-based model to accurately predict the batteries capacity fade over time, comparing projects that use second-life or new lithium-ion batteries. Our model allows us to assess the impact of control policies on long-term performance including limits to the depth-of-discharge (DoD), temperature control, and minimum and maximum state-of-charge (SOC) limits for storage systems. By setting certain control policy limits to minimize cycle aging, we show that a system with state-of-charge limits in a 65-15% range, extends the project life to over 16 years to improve project economics over those for a new battery that uses 85-20% limits. In comparison, control policies that more rapidly degrade the battery by using a large 80% DoD limit in a 95-15% range, to 'front-load' revenue during a projects life-cycle, were found to significantly underperform a new battery owing to the resulting shorter project life.

We extend our model to determine the conditions under which second-life batteries are more favorable to new ones. We show this can be supported by certain technical and economic conditions given here in order of importance: (i) utilizing charge control policies that minimize average SOC to extend battery life, (ii) second-life battery capital costs that are 60% or less than new ones, (iii) PV capex < 1000 $/kW (the current industry benchmark is 1111 $/kW [12]), and (iv) BOS costs of <150 $/kWh (the current industry benchmark is 171 $/kWh [12]).

In summary, our modeling shows that using current benchmarked data for the capital and O&M costs of photovoltaics plus energy storage systems, and a semi-empirical data-based degradation model, under certain conditions that include suitable physical end-of-life assumptions, PV and BOS capital cost conditions, second-life batteries offer a compelling alternative to new lithium-ion batteries in solar-plus-storage systems.

*Method Summary*

We model the performance of a combined photovoltaic power plant and energy storage system (PVESS) in California using 2017 price data from CAISO (oasis.caiso.com) and NREL's national solar radiation database (https://nsrdb.nrel.gov/) . The PVESS modeled is a DC-coupled system and we assume the battery can only be charged by the PV power directly, *i.e.*, the battery is never charged directly from the grid. We model two revenue streams for the PVESS project:

(i) Selling electricity on the merchant power market either directly from the PV system or storing the 4-hour battery system and selling from the battery. To optimize the revenue of the combined PVESS system, we use a function-minimizing optimization routine that takes a daily solar production curve and day-ahead market prices in CAISO, as shown in Figure 1, and optimizes the charge and discharge policy to maximize daily revenue following a method similar to [13]. The daily optimization routine does not consider any long-term degradation trade off.

(ii) The second revenue source is capacity credits, which we model based on the method in [14]. It was shown that PV projects typically obtain 40% of their rated capacity in credits for $149/kW each year. In addition, the full battery power is considered to contribute to the capacity credit, meaning our year 1 capacity credit for a new battery is $40,953. For second-life batteries and new batteries after year 1, we calculate the capacity credit based on the faded values at the start of each year, so the capacity credit reduces annually.

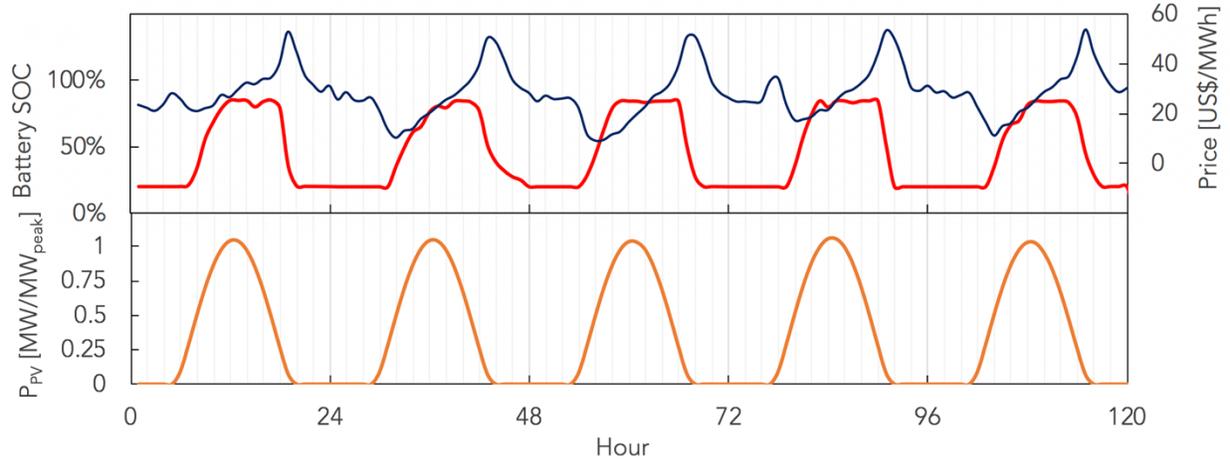

Figure 1: Sample of the power produced by the photovoltaic power plant (orange), the day-ahead spot market price (navy) and resulting optimized charge/discharge profile (red) of the energy storage system for 5 days.

At the end of each daily optimization routine, a data-based battery degradation method is used to calculate the incremental capacity fade. The battery capacity is updated accordingly for next day's operation. The battery degradation algorithm considers the non-linear growth of a solid-electrolyte interface early in a battery's life, calendar aging, and cycle aging. The calendar aging model considers the stress from the average battery state of charge over the daily 24-hour period, while the cycling aging model uses the rainflow counting algorithm to quantify cycling events. Both calendar and cycle aging models consider the effect of battery temperature, which we assume is maintained at 25ºC by the battery cooling system [9]. Our method does not try to predict the 'knee' in the battery capacity fade where the primary loss transitions to electrode-site loss, as opposed to lithium usage, and degradation accelerates. This behavior is expected to occur earlier in a battery's life for charge/discharge policies that allow high depths-of-discharge, but it has been shown that lithium-limited loss dominates in solar self-consumption scenarios [15].

The project costs consist of the initial capital costs including the cost to install the complete PV system, the cost of the battery module and the battery balance-of-system (BOS) costs, where we use the latest benchmarks from the US for utility-scale PVESS systems from [12] and presented in Table 1. A capital cost discount rate of 8.5% for all capital expenditures is assumed when constructing a DC-coupled system with both power and storage on the same site. Industry-benchmarked operations and maintenance costs are used and discounted over the project life.

We use the benefit-cost ratio as outlined in [14] to compare individual projects, it consists of the initial capital costs and discounted operations and maintenance costs, *O&M*, divided by the discounted revenue streams for each project as in Equation 1 where *N* is the project life in years and *r* is the discount rate:

$$Benefit\text{-}Cost\ Ratio = \frac{\sum_{i=0}^{N} \frac{Annual\ Revenue}{(1-r)^i}}{Initial\ capital\ costs + \sum_{i=0}^{N} \frac{O\&M\ costs}{(1-r)^i}} \quad (1)$$

It should be noted that in all our modelled cases, the project life is assumed to conclude once the battery reaches its physical end-of-life (EOL) at 70% or 60% of remaining useful capacity. In practice, the photovoltaic power system might be re-powered as a merchant solar power plant only or under a power-purchase agreement without the use of a battery, which would further extend the revenue generated by the project without additional capital expenditure and increase the benefit-cost ratio of the project. Additionally, battery systems might be replaced by new ones at their EOL. In the interest of simplicity, we do not attempt to model the costs and benefits of these cases but compare individual projects as a function of their battery life.

Table 1: System, cost and revenue assumptions used.

|  |  |  | Comment |
|---|---|---|---|
| PV Size | 2.5 | MW |  |
| Initial battery capacity | 10 | MWh | 4 hours |
| Initial battery power | 2.5 | MW |  |
| Charge/Discharge Efficiency | 90 | % |  |
|  |  |  |  |
| PV system | 1111 | $/kW (DC) | [12] |
| PV O&M | 11 | $/kW-year | [12] |
| New battery module | 209 | $/kWh | [12] |
| Battery BOS | 171 | $/kWh | [12] |
| Battery O&M | 9 | $/kW-year | [12] |
| DC-coupled discount | 8.5 | % | [12] |
| Capacity credit | 149 | $/kW-year | [14] |
| Discount rate | 7 | % | [16] |

Furthermore, we do not consider the impact of EOL costs. While some geographic jurisdictions have mandated the collection and recycling of lithium-ion batteries, this is not the case in the USA and we do not include these costs in this analysis. Further benefits might be found for second-life batteries in the ability to extend their usefulness by a number of years and delay the incurring of recycling costs for the responsible party, most likely the original electric vehicle manufacturer.

**Results**

We start by comparing the annual revenue of three possible projects; a 2.5 MW photovoltaic power plant only, a 2.5 MW photovoltaic power plant with a new 10 MWh lithium-ion NMC energy storage system, and a 2.5 MW photovoltaic power plant with a second-life 10 MWh lithium-ion NMC energy storage system consisting of batteries that have faded to 80% of their original capacity, *i.e.*, 8 MWh capacity is available on the first day of the project. The SOC of the battery systems is limited to between 85% and 20% of the available capacity and it is assumed the batteries are maintained at 25°C using a temperature control system. Figure 2 (a) shows the revenue versus year for the three projects including battery degradation and considering a discount rate of 7%. The project with a new NMC battery generates revenue of $669,000 in year 1 compared to $575,000 for the second-life battery, a difference of 16%. Figure 2 (b) outlines how both battery systems fade over time and reveal the drastic 8% reduction in capacity in the new battery during the first year of operation due to the non-linear effects of solid-electrolyte interface growth. In year 2, the revenues from both projects are $594,500 and $528,000 respectively, the difference between them reducing to 12.5% owing to the rapid early capacity fade of the new battery and the start of the influence of the discount rate.

After the initial SEI formation in the new battery, the capacity fading slows and in the following years fades at a rate of 2.6 – 1.5% per year. After year 1 the fade rates of both batteries are close to each other owing to their similar daily usage. After 10 years of operation, the new battery has faded from 100% to 75% useful capacity, a loss of 25%, while the second-life battery has faded from 80% to 63%, a loss of 17%. The greater loss for the new battery over this time period is due to the SEI film formation in year 1 causing a significant loss that is not present for the second-life battery.

The revenue from a stand-alone photovoltaic power plant is $282,500 in year 1, 42% and 49% of the revenue generated by the new and second-life storage projects, respectively. By year 10, given the capacity fade and discounting of revenue, the photovoltaic-only revenue as a percentage of the storage projects increases to 51% and 57%, respectively.%.

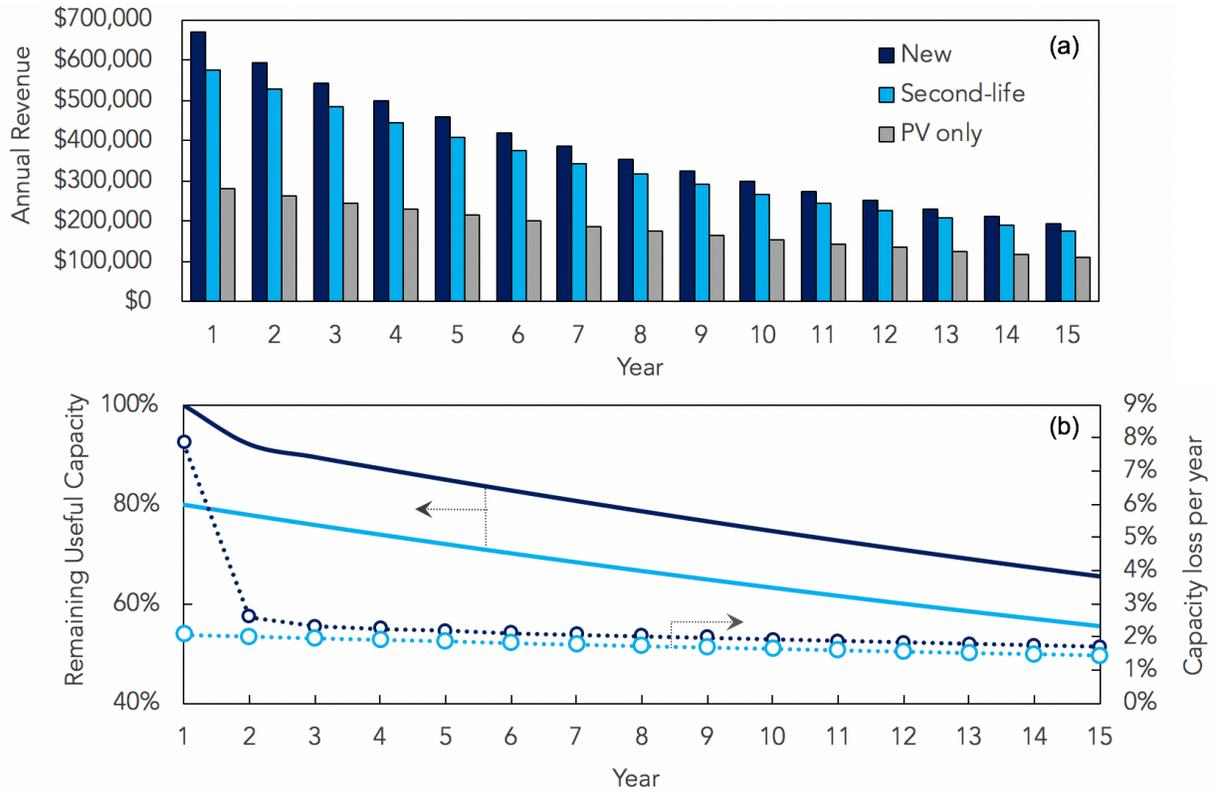

Figure 2: (a) The annual revenue of a stand-alone merchant photovoltaic power plant (PV only) and a photovoltaic + energy storage systems using new or second-life batteries, and (b) the remaining useful capacity of the batteries versus time and annual capacity fade for both energy storage projects modeled.

While the previous discussion analyzed the impact of battery capacity fade and the time value of money on potential revenues, it did not consider the battery end-of-life (EOL). In this section we begin to compare individual projects over their full life cycle by making assumptions of battery lifetimes. In order to justify the use of already faded batteries, we assume two cases are possible: lithium-ion batteries (new and second-life) can be used until their capacity reaches either 70% of their original capacity or 60%. The former is close to what is used for physical EOL in the literature [2], [4], while we assume the latter is probably required to produce second-life projects that last enough to justify the expense of deploying them. It should be noted that we continue using our linear model of battery degradation right to the battery breakpoint and do not consider additional degradation pathways that might be present in the later stages of a battery's life [15]. Our assumptions that a batteries EOL is at 70% or 60% is suitable for our purposes, but accurate prediction of this breakpoint for economic and safety reasons remains an active and important area of research [11].

To compare projects, we use a benefit-cost ratio for individual projects, as outlined in Equation 1, with our results presented in Table 2. This ratio divides the investment and discounted annual O&M costs by the projects discounted revenues over its life; larger values indicate more favorable project economics with values greater than 1 indicating profitable projects. For our analyses we model the project costs and revenues up until the battery reaches either 70% or 60% of its original capacity. For new batteries in our model, it takes 11.4 and 17.3 years to reach 70% remaining capacity while the second-life batteries take 5.1 and 11 years respectively – highlighting the need for accurate breakpoint prediction for financial projections where our assumption of an extra 10% available capacity fade results in more than a doubling of the expected life of the second-life battery. For modeling purposes, we assume the cost of the photovoltaic power plant, and energy storage balance of systems costs are the same for both projects, the new lithium-ion batteries are installed for $209/kWh while the second-life batteries cost half that amount, $104.5/kWh. With these assumptions, our project finance model shows a new battery system has a benefit-cost ratio of 0.76 or 0.93 for the 70% or 60% EOL cases. For reference, a 20-year photovoltaic-only project has a benefit-cost ratio of 1.04, indicating that given the current capital costs for lithium-ion batteries and likely revenue streams for the project modeled, photovoltaics-only remain the most attractive. Assuming a new battery lasts to 60% of its original capacity, allows for a 17.3-year project life that will almost generate enough revenue over this time to justify the initial expense.

The second-life battery projects, however, have lower benefit-cost ratios of 0.47 and 0.78 for the 70% and 60% EOL cases respectively. The low value of 0.47 shows how assuming the second-life battery project can only use 10% of the remaining capacity before EOL results in a very short project life that does not justify the capital expense. Assuming the second-life project will operate over 20% of the remaining useful life of the battery extends the project life to 11 years and results in significant project revenues of $428,206.

Table 2: The modeled costs, revenues, and benefit-cost ratios for PVESS projects considering a 70% and 60% end-of-life for an SOC limiting range of 85–20% and a discount rate of 7% where the second-life battery cost is assumed to be half of the new battery cost.

|  | EOL (Years) | Capital cost | O&M cost | Revenue | Benefit-cost ratio |
|---|---|---|---|---|---|
| New battery | | | | | |
| EOL 70% | 11.4 | $ 605,650.99 | $ 41,068.09 | $ 492,812.81 | 0.76 |
| EOL 60% | 17.3 | $ 605,650.99 | $ 52,708.10 | $ 609,731.23 | 0.93 |
| 2nd life battery (faded 80%, assumed 50% capital cost) | | | | | |
| EOL 70% | 5.1 | $ 509,428.22 | $ 22,292.55 | $ 248,357.38 | 0.47 |
| EOL 60% | 11 | $ 509,428.22 | $ 40,117.91 | $ 428,205.82 | 0.78 |

Figures 3 (a) & (b) show the range of benefit-cost ratios possible for the second-life battery projects as a function of the cost of the battery for the 60% and 70% EOL cases respectively. The 11-year lifetime of the second-life project is much less than the 17.3-years for new batteries for the 60% EOL case. A very lower battery price of 20% of the cost of a new battery (41.8 $/kWh) is required for the benefit-cost ratio to reach 0.87 and come close to the cost-benefit ratio of the new battery project of 0.93 that benefits from greater revenue from the larger battery as well as the revenue generated during the extra 6.3 years of the project's life. For the 70% EOL case, a similarly low second-life battery costs results in a benefit-cost ratio of 0.52 that is much less than the 0.76 value for a project with new batteries. These values highlight that using the same control policies for new and second-life batteries lead to more favorable project economics for projects with new batteries unless the second-life battery cost is close to zero.

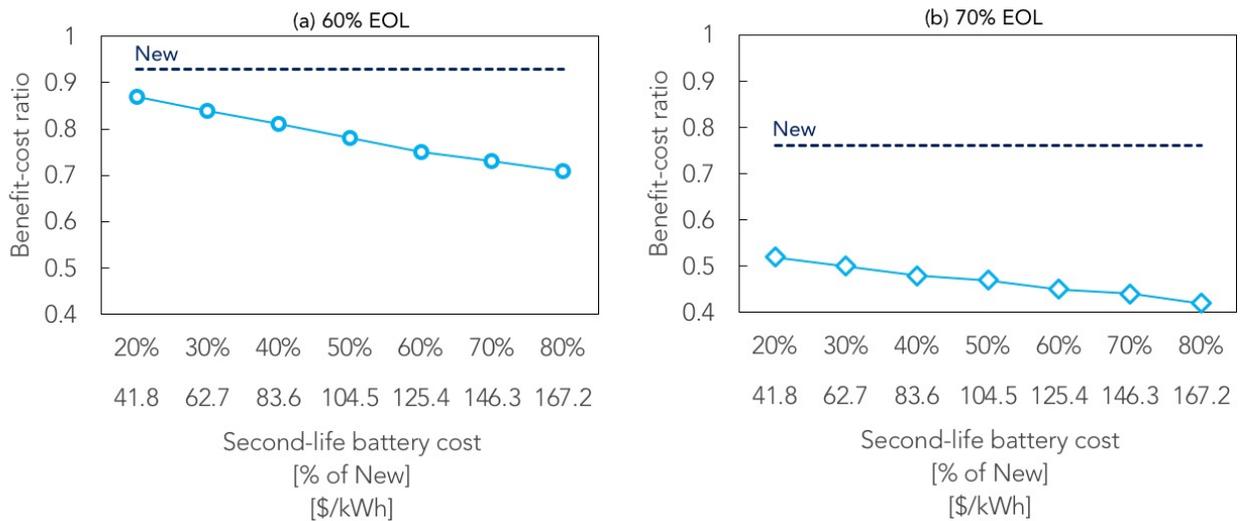

Figure 3: The benefit-cost ratio of the modeled second-life energy storage projects versus the capital cost to install the second-life batteries for the (a) 60% and (b) 70% EOL cases.

*The impact of control policy limits on second-life economics*

The use of data-based methods for battery degradation modeling allows us to trial different control policy limits and compare their impact on the long-term performance of a project. In our analyses so far, we have modeled the degradation of new and second-life batteries subject to a maximum depth-of-discharge (DoD) of 65% between maximum and minimum state-of-charge limits of 85% and 20%, while assuming the battery is maintained at 25°C throughout the project life. These limits are now altered to compare the impact of different policies on battery life. Specifically, we model two additional cases where; (i) a large DoD of 80% is used with 95% and 15% SOC limits respectively to analyze a case where the battery is rapidly degraded to 'front-load' revenue, and (ii) a smaller DoD of 50% is allowed between SOC maximum and minimums of 65% and 15% respectively, intentionally maintaining the average SOC at much lower values to minimize $SOC_{stress}$ degradation.

Figure 4 (a) shows the resulting capacity fade for these second-life control policy limits versus a new battery subject to the original 85–25% limits as modeled in the previous section. The 95–15% case leads to rapid degradation of the battery capacity to the 70% EOL limit in 3.9 years and the 60% EOL limit in 8.4 years. The corresponding benefit-cost ratios are 0.33 and 0.57 respectively for a second-life battery cost discount of 50% ($104.5/kWh) and a discount rate of 7%. The more conservative charge/discharge policy limits of 65–15% leads to much slower degradation, reaching 70% EOL in 7.5 years and 60% in 16.1 years, almost as long as the new battery system. The 60% EOL case achieves the same-benefit cost ratio as the new project of 0.93 when the second-life battery discount is assumed to be 50% with a 7% discount rate. Despite only using 50% of the available battery capacity at any time, by maintaining a low average SOC and reducing $SOC_{stress}$ degradation effects, this policy results in the highest benefit-cost ratios for the second-life storage projects and highlights how the best lever for increased financial performance is to extend the battery life as long as possible, even at the expense of revenue early in the project life.

Figures 4 (b) & (c) shows the benefit-cost ratios for projects that use these different control policies the versus the second-life battery cost. Both figures include the benefit-cost ratio of project with new batteries and it can be seen that for the 70% EOL case, no matter the second-life battery cost, a project with new batteries remains the most economically favorable. For the 60% EOL case, Figure 4 (b), the much longer project lifetime achieved by using a 65–15% control policy results in a benefit-cost ratio of 0.93, equal to the value for a project with new batteries, when the second-life battery cost is 80% of the new price or 167.2 $/kWh. Furthermore, this control policy leads to the first project design with a benefit-cost ratio greater than the break-even point. Given the assumptions already outlined, if EV manufacturers can collect, repurpose and supply used EV batteries for <125.4 $/kWh (<60% of current new prices), and control policies are implemented to extend battery life, profitable solar-plus-storage projects can be constructed for the location modeled by utilizing second-life batteries – something would not be possible with new batteries at current market prices.

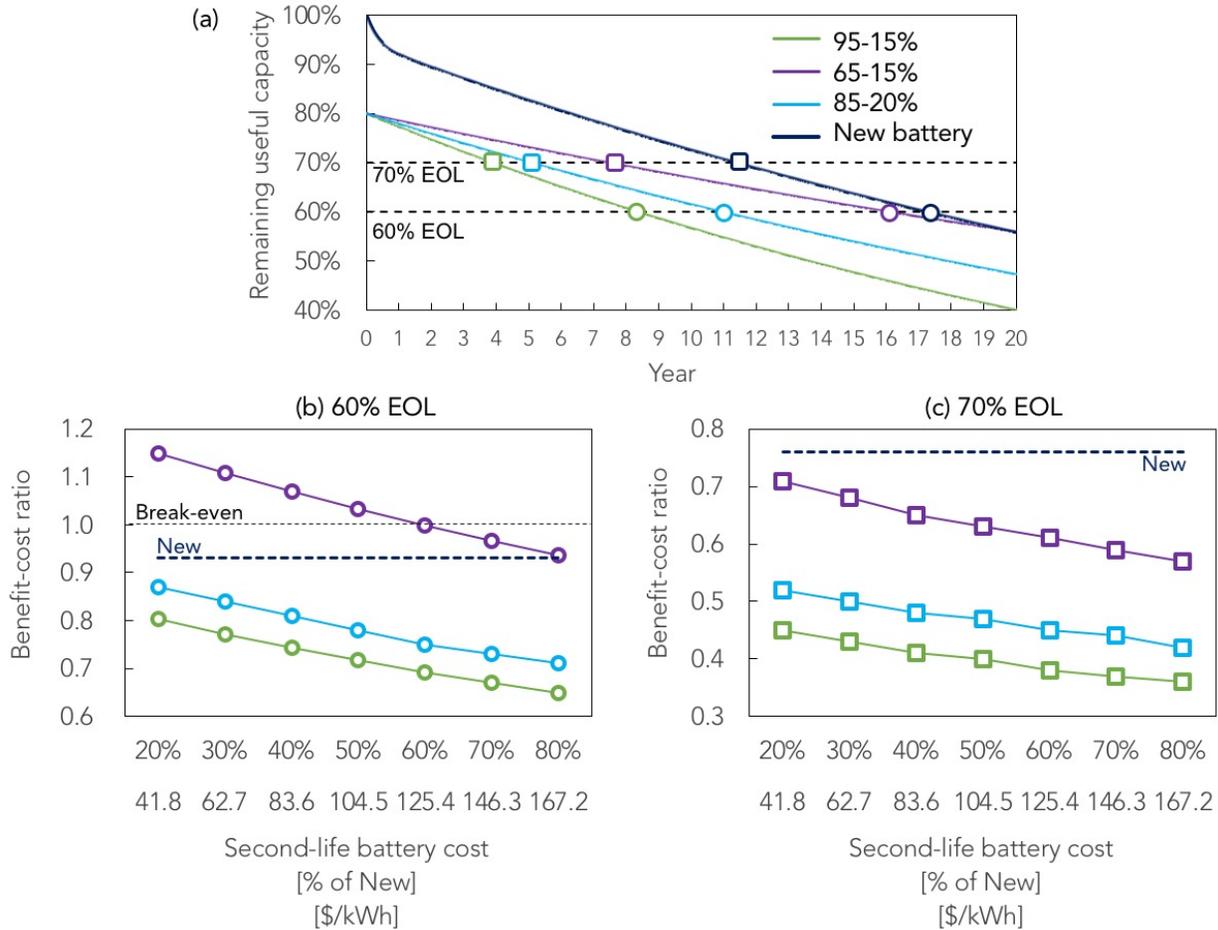

Figure 4: (a) The remaining useful capacity of the modeled batteries considering different control policies for the second-life batteries where open squares mark the 70% EOL and open circles mark the 60% EOL, (b) the benefit-cost ratio for each control policy modeled versus the capital cost to install the second-life batteries for the 60% EOL assumption, and (c) for the 70% EOL assumption.

Considering the difference between control policies, the use of the rainflow counting algorithm allows us to visualize cycling behavior of both policies as relevant to fatigue analysis. Figure 5 (a) & (b) compare the 20-year cycling behavior of the modeled SOC limit policies of 65–15% and 95–15%, respectively. For the 95–15% limits, the second-life batteries undergo 1000's of cycles in a 300–600 kWh range around the battery mid-point. The 65–15% limits constrain the majority of cycles to within a 300–400 kWh range, showing the impact of limiting the maximum DoD and average SOC over the lifetime of the battery's use significantly reduces cycle aging.

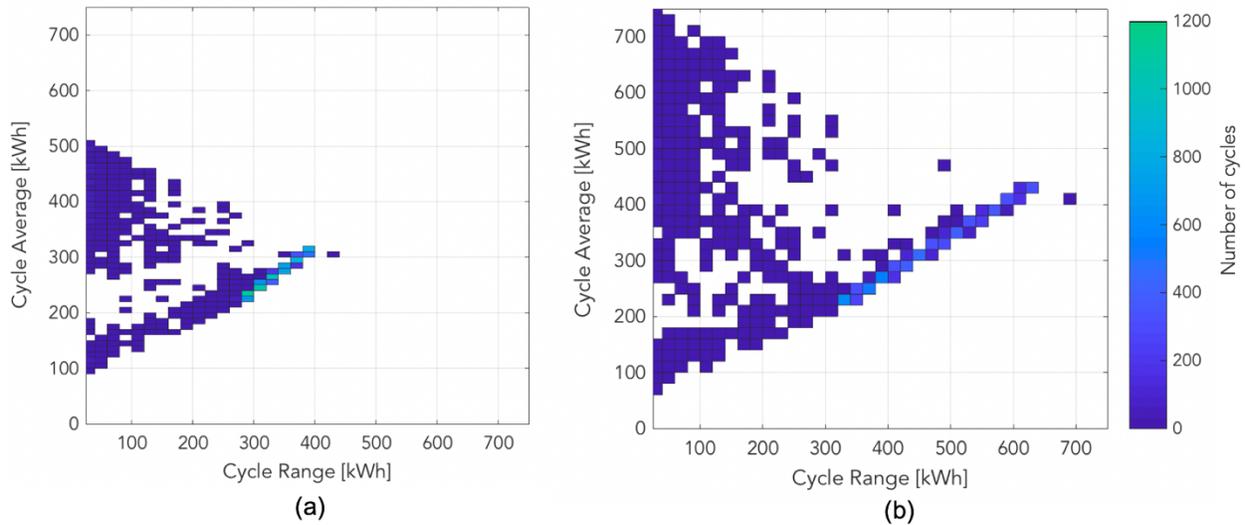

Figure 5: The number of cycles binned by cycle range and cycle average modeled over 20 years of a project life for (a) a 65–15% and (b) a 95–15% SOC limit control policies, respectively.

*Sensitivity analysis for cost and revenue assumptions*

Based on the 2018 benchmark costs for solar and storage systems, the benefit-cost ratio for second-life solar-plus-storage systems we model are typically less than the values for new batteries, except for cases where a low 65-15% SOC limit is implemented. In this section, we perform a sensitivity analysis of a number of important costs and revenue streams to quantify which cost declines or revenue increases would have a greater impact on second-life battery projects versus new ones. Given the 65-15% policy and an assumed EOL of 60% results in the highest benefit-cost ratios, we use these results here. Fig. 6 outlines the results of three sensitivity analysis where the balance of systems costs [$/kWh], photovoltaic capital cost [$/kW] and the capacity credit [$/kW] are all varied. Fig. 6 (a) shows that reducing the balance of systems costs from 200 to 100 $/kWh for a new project results in the benefit-cost ratio increasing from 0.89 to >1. Fig. 6 (b) outlines the benefit cost ratio for a second-life project where the balance of system costs are varied over the same range while a range of second-life battery cost fractions from 0.2 – 0.8 of the original capital expense are considered. As can be seen, benefit-cost ratios of >1 are possible for BOS values of <180 $/kWh when combined with low second-life battery costs. The most interesting values however are the difference between the benefit-cost ratio for a new versus a second life project when the BOS reduction is applied to both. Fig. 6 (c) shows the difference between the benefit-cost ratio of the second-life and new project, where positive values indicate the second-life project is more worthwhile. We see that there is a significant difference between the benefit-cost ratios as the BOS and the second-life fraction reduce. Figs. 6 (d)-(f) summarize the same analysis where the photovoltaic capital cost is varied from 100 – 200 $/kW.

Again, these reductions have a significant impact on which project type is more profitable with PV costs of <1000 $/kW and second-life battery cost fractions of 0.5 and less resulting in benefit-cost ratio differences of 0.1-0.3.

Lastly, Figs. 6 (g)-(i) outlines the results for an analysis where the capacity credit is varied from 100 – 200 $/kW. In this case, the increase in capacity credit has a significant and similar impact on both project types and does not benefit the second-life project enough to make them more profitable than a new project.

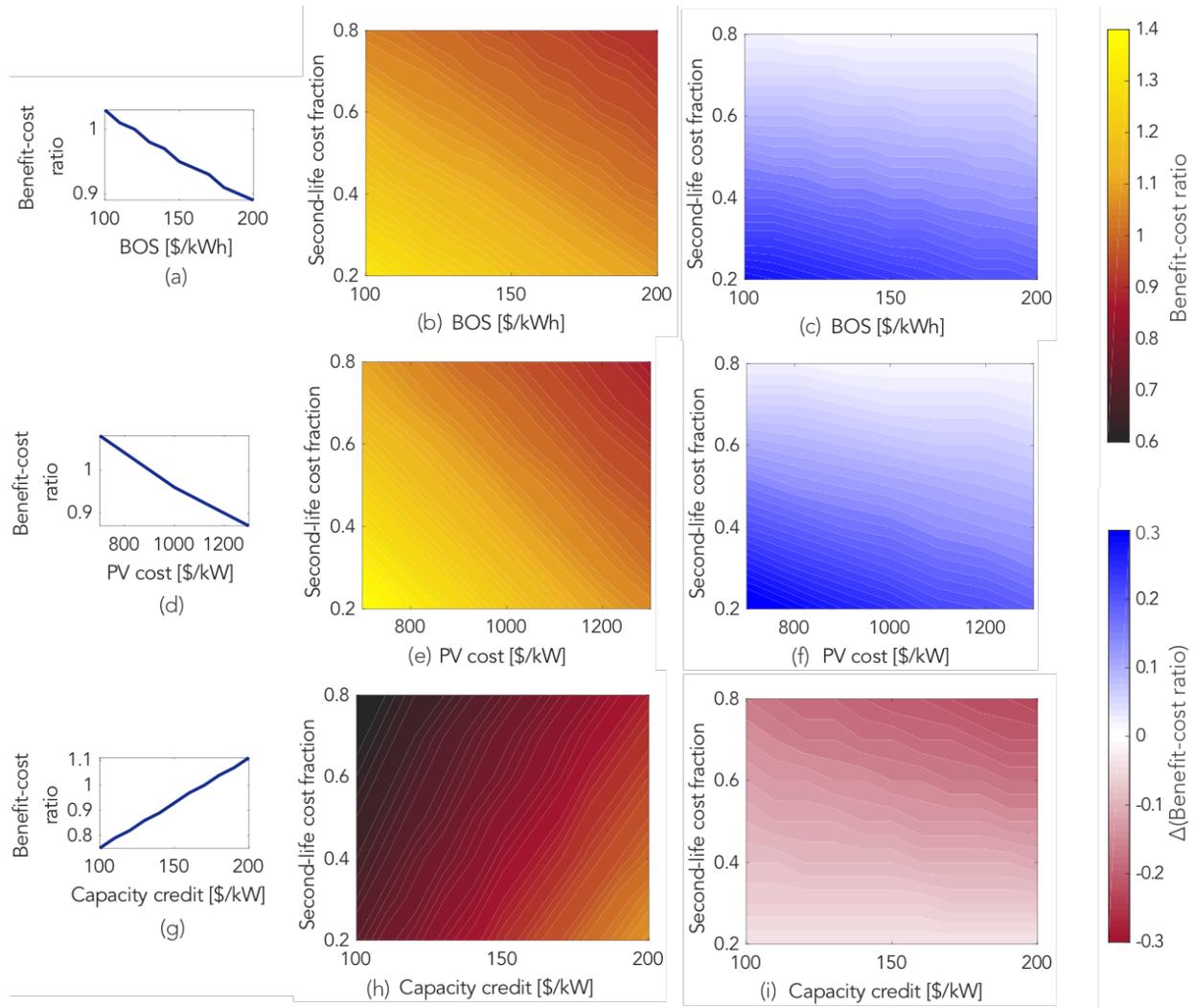

Figure 6: The benefit-cost ration for a solar-plus-storage project for (a) new batteries versus BOS costs, (b) second-life batteries versus battery and BOS costs, (c) the difference between new and second-life benefit-cost ratios versus second-life battery and BOS costs, (d) new batteries versus PV cost, (e) second-life batteries versus battery and PV costs, (f) the difference between new and second-life benefit-cost ratios versus second-life battery and PV costs, (g) new batteries versus capacity credit, (h) second-life batteries versus battery cost and capacity credit and, (i) the difference between new and second-life benefit-cost ratios versus second-life batter costs and capacity credit.

**Conclusions**

The use of stationary energy storage systems combined with photovoltaic power plants is growing to mitigate impacts of the variability of the solar resource on the grid. Given current lithium-ion battery prices, there remains uncertainty about the profitability of combined solar power and energy storage systems. At the same time, the rapid proliferation of electric vehicles is creating a fleet of millions of lithium-ion batteries that will be deemed unsuitable for the transportation industry once they reach 80% of their original capacity. The repurposing and deployment of these batteries as stationary energy storage provides an opportunity to reduce the cost of storage for multiple applications if the economics of using old batteries can be proven.

In this paper, we modeled the economic performance of a combined photovoltaics plus second-life energy storage project in California including a data-driven, semi-empirical model of lithium nickel manganese cobalt oxide battery degradation to predict its capacity fade over time, and compared it to a project that used new lithium-ion batteries. By setting certain control policy limits, to minimize cycle aging, we show that a system with SOC limits in a 65–15% range, extends the project life to over 16 years. Under these conditions, a second-life project is more economically favorable than a project that uses a new battery and 85–20% SOC limits, for second-life battery costs that are <80% of the new battery. The same system reaches break-even and profitability for second-life battery costs that are <60% of the new battery. In comparison, control policies that more rapidly degrade the battery by using a large 80% DoD limit in a 95–15% range, to 'front-load' revenue during a project life-cycle, were found to significantly underperform a new battery owing to the resulting shorter project life.

Our model shows that using current benchmarked data for the capital and O&M costs of solar-plus-storage systems, and a semi-empirical data-based degradation model, it is possible for EV manufacturers to sell second-life batteries for <60% of their original price to solar-plus-storage projects developers. In summary, we have shown there is significant value remaining in used EV batteries and in solar-plus-second-life projects provided a strong incentive to bring together the multiple stakeholders required to build this technology at scale.

**Acknowledgements**

The authors acknowledge the sources of funding for this work. I.M. has received funding from the European Union's Horizon 2020 research and innovation programme under the Marie Skłodowska Curie grant agreement No. 746516. I.M.P. was financially supported by the DOE-NSF ERF for Quantum Energy and Sustainable Solar Technologies (QESST) and by funding from Singapore's National Research Foundation

**Methods**

*Optimizing daily battery charging and discharging to maximize revenue*

The per-hour charge and discharge cycles of the energy storage system were optimized on a daily basis to maximize the revenue from the complete PVESS system. To simulate the operation of a PVESS system, we first assume the operator knows the wholesale market electricity price and solar resource availability at the site 24-hours in advance, and maximize the revenue possible using a constrained nonlinear multivariable function subject to certain constraints. The PV system can either sell power directly to the grid as it is produced, or store all or a portion of the available energy in the battery to discharge and sell at a later time. The optimization was subject to a number of constraints, namely: the battery could only be charged from available PV power and not directly from the grid, maximum battery charge and discharge rates were set equal to the battery's power, the battery state-of-charge (SOC) could not exceed or go below maximum and minimum limits and, the process of charging or discharging of the battery had an efficiency of 90%. As the capacity of a modeled battery faded over time, the SOC limits accordingly adjusted accordingly to the available capacity. The optimization routine is outlined below where; $P_{PV}$ (MW) is the power produced by the photovoltaic power plant, $t$ (hour) is time, $x$ (MW) is the charge/discharge power of the battery, *price* (US$/MWh) is the day-ahead market price, $C_{max}$ and $C_{min}$ (MW) are the charge and discharge limits and $SOC_{max}$ and $SOC_{min}$ (MWh) are the state of charge limits.

When calculating revenue, two case are considered:

(i) Solar power is available and revenue can be made by selling to the grid or the battery can be charged for later use:

$$Revenue_{solar}(\$) = \sum_{t=0}^{t=24} -((P_{PV}(t) - x(t)) * price(t))$$

(ii) Solar power is not available and the only revenue is derived by discharging the battery:

$$Revenue_{battery}(\$) = \sum_{t=0}^{t=24} x(t) * price(t)$$

The optimized control policy for a 24-hour period maximizes the sum of these revenues:

$$Revenue_{total}(\$) = max(Revenue_{solar} + Revenue_{battery})$$

subject to:

$$C_{discharge}^{min} \leq x \leq C_{charge}^{max}$$

$$x \leq P_{PV}$$

$$SOC(t) = SOC(t-1) + x(t)$$

$$SOC_{min} \leq SOC \leq SOC_{max}$$

*Lithium-ion battery degradation using a data-based model*

After each 24-hour period, the battery capacity is degraded in line with its use. The degradation behavior of lithium-ion batteries is impacted by many factors including the temperature at which the battery is maintained, the average state-of-charge (SOC) maintained in the battery, the depth and frequency of cycling as well as calendar aging. We use a data-based semi-empirical lithium nickel manganese cobalt oxide battery degradation model that assesses battery cell life loss from operating profiles as summarized here and described in detail in [9].

Non-linear capacity fade of a new battery, *L*, is given below where *f* is the linear cycle and calendar aging and $A_{SEI}$ and $B_{SEI}$ are fitted empirical parameters with the values for NMC batteries given in Table 3:

$$L = 1 - \alpha_{sei} e^{-\beta_{sei} f} - (1 - \alpha_{SEI})e^{-f}$$

Table 3: SEI parameters used for NMC batteries

| | |
|---|---|
| $A_{SEI}$ | 0.0575 |
| $B_{SEI}$ | 121 |

For batteries that have already undergone SEI film formation and now undergo linear degradation only, as for our second-life batteries, we use a simplified linear equation where $L'$ is the prior capacity fade (= 0.2 at the start of our second-life battery project model):

$$L = 1 - (1 - L')e^{-f}$$

Capacity fade is a function of calendar and cycling aging. Calendar aging begins from the start of the batteries life and is caused by the loss of cyclable lithium. Cycle aging occurs when the battery is used and is typically caused by active material structure degradation and mechanical fracture.

To quantify the irregular cycling patterns of the battery into the relevant parameters required to predict battery degradation, we start by using a rainflow cycle-counting algorithm as is widely used in fatigue studies. We input the SOC over 24 hours and the algorithm returns whether each cycle is a full or half-cycle, the cycle depth, the cycle mid-point, cycle begin and end times and the total number of cycles.

In our model, cycles and time are modeled as actual factors that reduce the life of the battery, while SOC and temperature influence the rate of degradation:

$$f = [S_\delta(\delta) + S_t(t)]S_\sigma(\sigma)S_T(T_c)$$

Where $S_\delta$ is the depth-of-discharge stress modeled as a quadratic for NMC batteries = $k_{\delta 1}\delta^{k_{\delta 2}}$, where $k_{\delta 1} = 0.2/(3000*0.8^{k_{\delta 2}})$, representing a cycle life of 3000 cycles at 80% DoD, until 80% end of life, and $k_{\delta 2}$ is an empirical non-linear coefficient of 2.03.

The calendar aging stress, $S_t = k_t t$, where $t$ is the battery age in seconds and $k_t = 4.1375 \times 10^{-10}$/s.

The models use reference points for SOC, $\sigma_{\text{ref}} = 0.6$, and temperature, $T_{\text{ref}} = 25°C$, at which the stress model has a value of one, indicating that the degradation rates are unaffected at the reference condition.

$S_\sigma$ is the SOC stress where $\sigma$ is the average SOC during a cycle and $\sigma_{ref}$ is the reference SOC value.

$$S_\sigma = e^{k_\sigma(\sigma - \sigma_{\text{ref}})}$$

$S_T$ is the temperature stress based on the Arrhenius equation:

$$S_T = e^{k_T(T - T_{\text{ref}}) \cdot \frac{T_{\text{ref}}}{T}}$$

*The code for this lithium nickel manganese cobalt oxide battery degradation model is available at https://bolunxu.github.io/*